\documentclass[11pt,oneside,onecolumn,draftcls]{IEEEtran}
\usepackage[dvips]{graphicx}
\usepackage{epsfig}
\usepackage{amsmath}
\usepackage{epsfig}
\usepackage{array}
\usepackage{amssymb}
\usepackage{color}

\newtheorem{Theorem}{\textbf{Theorem}}

\begin{document}

\title{A Simple Method for Secret-Key Generation Between Mobile Users Across Networks}

\author{Yingbo Hua\thanks{Department of Electrical and Computer Engineering,
University of California at Riverside, Riverside, CA 92521, USA. Email: yhua@ece.ucr.edu. This work was supported in part by the U.S. Department of Defense under W911NF-20-2-0267. The views and conclusions contained in this
document are those of the author and should not be interpreted as representing the official policies, either
expressed or implied, of  the U.S. Government. The U.S. Government is
authorized to reproduce and distribute reprints for Government purposes notwithstanding any copyright
notation herein.}
}

\maketitle
\begin{abstract}
Two or more mobiles users can continuously superimpose sequences of bits chosen from different packets or files already exchanged and authenticated between themselves to continuously renew a secret key for continuous strengthening of their privacy and authentication. This  accumulative, adaptable and additive (AAA) method is discussed in this paper.  The equivocation to Eve of any bit in the generated key by the AAA method equals to the probability that not all corresponding independent bits exchanged between the users are intercepted by Eve. This performance, achieved without using any knowledge of non-stationary probabilities of bits being intercepted by Eve, is compared to an established capacity achievable using that knowledge.  A secrecy robustness of the AAA method against some correlations known to Eve is also discussed.
\end{abstract}

\section{Introduction}
Secret-key generation has been for decades a research topic of great interest. The fundamentals and extensive surveys of this topic are available in \cite{Bloch2011}, \cite{Poor2017} and \cite{Zhang2020}. Some latest developments on this topic are shown in \cite{HessamMahdavifar2020}, \cite{Li2022}, \cite{Ji2022}, \cite{MaksudHua2022}, \cite{HuaSP2023}, \cite{Hua2023April}, \cite{Hua2023Sept}, \cite{HuaMaksud2024}, and the references therein. These works are mostly focused on physical-layer key generation where the physical-layer channel reciprocity and/or noises  are exploited. Important issues for the physical layer approach still remain, and further research is clearly warranted. But for vast application scenarios in modern networks, reliable and authenticated connections between users across networks are already available. How to take advantage of this reality to build a strong secret key between remote users (without a dedicated physical channel between them) is clearly a question that should be answered. Among many benefits, such a secret key can be used to strengthen confidential and authenticated connections at a level higher than that based on public (and private) keys managed by a certificate authority or a central server.

This paper attempts to answer the above question by examining a simple method where each user continuously superimposes sequences of bits chosen from packets or files already exchanged and authenticated between themselves to continuously renew their secret key. The secrecy performance of this accumulative, adaptable and additive (AAA) method is presented, and compared with an established secrecy capacity achievable with additional conditions. Also discussed is a secrecy robustness of the AAA method against some dependency among the chosen bits.

\section{The Method and Its Performance}
Consider mobile networks in which two or more (legitimate) users can exchange data packets reliably. This is despite potential interferences and/or jamming where additional resources such as power and bandwidth are generally needed to achieve reliable transmissions within a given time window. To ensure the highest level privacy of their communications, for example, they need to have their own secret key to  encrypt and decrypt their transmitted packets. The conventional public-key based schemes can provide strong computation-based security (as opposed to information-theoretic security) but are costly in computation. They are typically implemented via central servers  for transport layer security \cite{Katz_Lindell_2021}, but are not easily implementable for peer to peer applications at the application layer. Furthermore, the secrecy of any given key in general weakens each time it has been used. This paper presents a simple method for mobile users (often remote from each other) to dynamically generate and continuously enhance a secret key between themselves  especially at the application layer.

Without loss of generality, let $X_{l,i}$ be the $l$th bit chosen (or transformed via any suitable function) from a packet or file already exchanged between the users at time $i$, where $l=1,\cdots, L$ and $i=1,2,\cdots$. For a given $i$, $X_{l,i}$ for different $l$ may or may not be chosen from the same packet or file. For a given $l$, $X_{l,i}$ for different $i$ should be chosen from packets or files exchanged at different times. So, $i$ is a time index while $l$ is like a spatial index.

The observed bit $E_{l,i}$ by Eve, corresponding to $X_{l,i}$, is either equal to $X_{l,i}$ (i.e., intercepted by Eve) or absent denoted by $\dagger$ (i.e., missed by Eve or an erasure for Eve). We will let $1-\mu_{l,i}=\texttt{Prob}\{E_{l,i}=X_{l,i}\}$ and hence $\mu_{l,i}=\texttt{Prob}\{E_{l,i}=\dagger\}$. For notational convenience, we will also use $x_{l,i}$ for both the random variable $X_{l,i}$ and its realization, and similar notations are also used for other variables.

The $l$th bit in an $L$-bit key generated by the AAA method at time $j$ is defined to be
\begin{equation}\label{eq:Kj}
  K_{l,j} \doteq X_{l,1}\oplus \cdots\oplus X_{l,j}
  =K_{l,j-1}\oplus X_{l,j}
\end{equation}
 with $\oplus$ denoting the binary sum (i.e., exclusive OR). Furthermore, $K_{l,i}$ at any time $i$ for all $l$ can be re-initialized if needed. Clearly, \eqref{eq:Kj} represents a continuous (or successive) privacy amplification, which compresses $X_{l,1}, \cdots, X_{l,j}$ into a single bit $K_{l,j}$ without requiring any knowledge of $\mu_{l,1},\cdots,\mu_{l,j}$.

 While the simplicity of the AAA method is clear and attractive, we now need to address its secrecy performance. Furthermore, we will also compare the AAA method with a capacity-achieving method in section \ref{sec:discussion}.

 \subsection{Performance of the AAA Method}

The secrecy performance of the AAA method at time $j$ can be measured by the following equivocation of the key $\mathcal{K}_j^{(L)}=\{K_{1,j},\cdots,K_{L,j}\}$ to Eve:
 \begin{equation}\label{eq:def_new}
   \epsilon_j^{(L)} \doteq \mathbb{H}(\mathcal{K}_j^{(L)}|\mathcal{E}_j^{(L)}),
 \end{equation}
 which is the entropy of $\mathcal{K}_j^{(L)}$ conditioned on $\mathcal{E}_j^{(L)}=\{E_{l,i};\forall l=1,\cdots,L; \forall i=1,\cdots,j\}$.

\begin{Theorem}\label{Theorem}
If $\{X_{l,i},E_{l,i}\}$ for each and every $l$ and $i$ is an independent realization of a random process, then $\epsilon_j^{(L)} = \sum_{l=1}^L\epsilon_{l,j}$ with
  \begin{equation}\label{eq:Ej}
    \epsilon_{l,j} \doteq\mathbb{E}\{K_{l,j}|\mathcal{E}_{l,j}\}= 1-\prod_{i=1}^j(1-\mu_{l,i})
  \end{equation}
  with $\mathcal{E}_{l,j}=\{E_{l,1},\cdots,E_{l,j}\}$.
It follows from \eqref{eq:Ej} that
\begin{enumerate}
\item $\epsilon_{l,j}\geq \epsilon_{l,j-1}$, i.e., $\epsilon_{l,j}$ is a non-decreasing function of $j$.
  \item $\epsilon_{l,j}=1$  (i.e., complete secrecy against Eve) for all $j\geq j_0$ if there is $i\in\{1,\cdots,j_0\}$ such that $\mu_{l,i}=1$.
  \item If $0<\mu_{l,i}<1$ for $i\in\mathcal{S}\subset\{1,\cdots,j\}$, then $\epsilon_{l,j}\to 1$ as $|\mathcal{S}|\to\infty$ where $|\mathcal{S}|$ is the size of the set $\mathcal{S}$.
  \item $\epsilon_{l,j}=0$ if and only if $\mu_{l,i}=0$ for all $i\in \{1,\cdots,j\}$.
\end{enumerate}
\end{Theorem}
\begin{IEEEproof}
It is obvious that $\epsilon_j^{(L)} = \sum_{l=1}^L\epsilon_{l,j}$ follows from the independence assumption. The statements following \eqref{eq:Ej} are also easy to verify.
So, we only need to prove \eqref{eq:Ej}.

Since we now focus on an arbitrary $l$, we can drop the index $l$, and replace $X_{l,i}$, $E_{l,i}$ and $K_{l,i}$ by $X_i$, $E_i$ and $K_i$. Also define $\mathcal{E}_j=\{E_1,\cdots,E_j\}$.
Let the probability distribution function of $K_j$ conditioned on $\mathcal{E}_j$ be denoted by $p(k_j|\mathbf{e}_j)$ with $\mathbf{e}_j=(e_1,\cdots,e_j)$.   Similar notations are defined accordingly.

It follows from $\epsilon_j \doteq\mathbb{E}\{K_j|\mathcal{E}_j\}$ (same as $\epsilon_{l,j}$ with $l$ dropped) that
\begin{equation}\label{eq:Hkjzj}
  \epsilon_j
  =-\sum_{k_j,\mathbf{e}_j} p(\mathbf{e}_j)p(k_j|\mathbf{e}_j)\log p(k_j|\mathbf{e}_j).
\end{equation}
The logarithm $\log$ has the base two.
First, consider $j=1$, for which
\begin{equation}\label{}
  p(k_1|\mathbf{e}_1)=p(x_1|e_1)
  =p(e_1|x_1)p(x_1)/p(e_1)
\end{equation}
with $p(e_1)=\sum_{x_1}p(e_1|x_1)p(x_1)$. The $x_1$ (row) vs $e_1$ (column) matrix of $p(e_1|x_1)$ is
\begin{equation}\label{eq:pe1x1}
  \begin{array}{c|ccc}
     & 0 & 1 & \dagger\\
    \hline
    0 & 1-\mu_1 & 0 & \mu_1 \\
    1 & 0 & 1-\mu_1 & \mu_1
  \end{array}
\end{equation}
and hence the distribution of $p(e_1)$ is
\begin{equation}\label{eq:pe1}
  \begin{array}{c|ccc}
     & 0 & 1 & \dagger \\
     \hline
     p(e_1)& \frac{1}{2}(1-\mu_1) & \frac{1}{2}(1-\mu_1) & \mu_1
  \end{array}
\end{equation}
Then \eqref{eq:Hkjzj} reduces to
\begin{align}\label{eq:E1}
  &\epsilon_1
  =-\sum_{k_1,\mathbf{e}_1} p(\mathbf{e}_1)p(k_1|\mathbf{e}_1)\log p(k_1|\mathbf{e}_1)=-\sum_{x_1,e_1} p(e_1)p(x_1|e_1)\log p(x_1|e_1)\notag\\
  &=-\sum_{x_1,e_1} p(x_1)p(e_1|x_1)\log [p(e_1|x_1)p(x_1)/p(e_1)]=\mu_1,
\end{align}
where the last equality follows from the substitutions of \eqref{eq:pe1x1} and \eqref{eq:pe1}.

Now consider $j> 1$.
We can write
$p(k_{j+1}|\mathbf{e}_{j+1})=p(k_j\oplus x_{j+1}|\mathbf{e}_j,e_{j+1})$, which equals to
$p(k_j|\mathbf{e}_j),p(k_j\oplus 1|\mathbf{e}_j),\frac{1}{2}$ for $e_{j+1}=0,1,\dagger$, respectively.
Note that if $e_{j+1}$ is 0 (or 1), so is $x_{j+1}$, and that if $e_{j+1}=\dagger$, $K_{j+1}=K_j\oplus X_{j+1}$  is a uniform binary random variable (due to the uniform binary random $X_{j+1}$).

We  know that the distribution $p(e_{j+1})$ is the same as \eqref{eq:pe1} after $\mu_1$ is replaced by $\mu_{j+1}$. Also $p(e_{j+1}=x_{j+1})=\frac{1}{2}(1-\mu_{j+1})$ for each of $x_{j+1}\in \{0,1\}$.

Therefore, it follows from \eqref{eq:Hkjzj} that, using $\mu_i' \doteq 1-\mu_i$ and $\bar k_i\doteq k_i\oplus 1$,
\begin{align}\label{eq:Hkjzj2}
  &\epsilon_{j+1}
  =-\sum_{k_{j+1},\mathbf{e}_{j+1}} p(\mathbf{e}_{j+1})p(k_{j+1}|\mathbf{e}_{j+1})\log p(k_{j+1}|\mathbf{e}_{j+1})
  =-\frac{1}{2}\mu'_{j+1}
  \sum_{k_j,\mathbf{e}_j} p(\mathbf{e}_j)p(k_j|\mathbf{e}_j)\log p(k_j|\mathbf{e}_j)\notag\\
  &\,\,-\frac{1}{2} \mu'_{j+1}
  \sum_{k_j,\mathbf{e}_j} p(\mathbf{e}_j)p(\bar k_j|\mathbf{e}_j)\log p(\bar k_j|\mathbf{e}_j)+\mu_{j+1}\sum_{k_{j+1},\mathbf{e}_j}p(\mathbf{e}_j)\frac{1}{2}\log 2\notag\\
  &=(1-\mu_{j+1})\epsilon_j+\mu_{j+1}.
\end{align}
Note that $\epsilon_j=-\sum_{k_j,\mathbf{e}_j} p(\mathbf{e}_j)p(k_j|\mathbf{e}_j)\log p(k_j|\mathbf{e}_j)
=-\sum_{k_j,\mathbf{e}_j} p(\mathbf{e}_j)p(\bar k_j|\mathbf{e}_j)\log p(\bar k_j|\mathbf{e}_j)$, and $\sum_{\mathbf{e}_j}p(\mathbf{e}_j)=1$.

Furthermore, we can now assume $\epsilon_j=1-\prod_{i=1}^j(1-\mu_i)$ since it is true for $j=1$ (see \eqref{eq:E1}). Then one can verify from \eqref{eq:Hkjzj2} that
\begin{equation}\label{eq:Ej1}
  \epsilon_{j+1}
  =1-\prod_{i=1}^{j+1}(1-\mu_i).
\end{equation}
Therefore, by induction, \eqref{eq:Ej1} holds for all $j\geq 0$. The proof is completed.
\end{IEEEproof}

An alternative proof of \eqref{eq:Ej}  is outlined here. First, one can verify that if any of the independent uniform binary variables $X_{l,1},\cdots,X_{l,j}$ is not intercepted by Eve, then $K_{l,j}$ in \eqref{eq:Kj} is a uniform binary random number with or without condition on $\mathcal{E}_{l,j}$, which is equivalent to $\mathcal{E}_{l,j}$ being independent of $K_{l,j}$. Second, if all of  $X_{l,1},\cdots,X_{l,j}$ are intercepted by Eve, then $K_{l,j}$ is known to Eve. Let $p_0$ be the probability that $\mathcal{E}_{l,j}$ is independent of $K_{l,j}$, and hence $1-p_0$ be the probability that $\mathcal{E}_{l,j}$ determines $K_{l,j}$. One can then verify (similar to \eqref{eq:E1}) that $\mathbb{H}\{K_{l,j}|\mathcal{E}_{l,j}\}=p_0$. Finally, it is obvious that $p_0$ equals to the probability that not all of $X_{l,1},\cdots,X_{l,j}$ are intercepted by Eve, which is \eqref{eq:Ej}.

\begin{Theorem}\label{Theorem_2}
If $X_{l,i}$ for all $l=1,\cdots,L$ are binary bits chosen randomly from file $i$, then
$\epsilon_j^{(L)}=L\left (1-\prod_{i=1}^j(1-\mu_i)\right )$ where $1-\mu_i$ is the probability that file $i$ is intercepted by Eve. (The intercept process for each file is assumed to be independent for all other files.)
\end{Theorem}
\begin{IEEEproof}
First let $L=2$, $\mathcal{X}_{1,j}=\{X_{1,1},\cdots,X_{1,j}\}$ and $\mathcal{X}_{2,j}=\{X_{2,1},\cdots,X_{2,j}\}$. The observations by Eve corresponding to $\mathcal{X}_{1,j}$ and $\mathcal{X}_{1,j}$ are denoted by $\mathcal{E}_{1,j}$ and $\mathcal{E}_{2,j}$. In this case, $\{\mathcal{X}_{1, j},\mathcal{E}_{1,j}\}$ is not independent of $\{\mathcal{X}_{2, j},\mathcal{E}_{2,j}\}$ because  $\{X_{1, i},E_{1,i}\}$ is not independent of $\{X_{2, i},E_{2,i}\}$ (unlike that in Theorem \ref{Theorem}). In fact, if file $i$ is intercepted by Eve, then $X_{1,i}$
and $X_{2,i}$ are constants when conditioned on  $\{\mathcal{E}_{1,j},\mathcal{E}_{2,j}\}$. (We assume that Eve knows all the protocols for the selection of  $X_{l,i}$ from file $i$.) But if file $i$ is missed by Eve, $X_{1,i}$
and $X_{2,i}$ are independent uniform variables (from Eve's perspective) with or without the condition on  $\{\mathcal{E}_{1,j},\mathcal{E}_{2,j}\}$.

Therefore, if all of the $j$ files are intercepted by Eve, the two bits  $K_{1,j}$ and $K_{2,j}$ in the key are fully exposed to Eve; but if any of the $j$ files is missed by Eve, $K_{1,j}$ and $K_{2,j}$ are independent uniform binary variables with or without the condition on $\{\mathcal{E}_{1,j},\mathcal{E}_{2,j}\}$. Using the above reasoning for any $L$ (as well as some elements in the proof for Theorem \ref{Theorem}) completes the proof.
\end{IEEEproof}

Theorem \ref{Theorem_2} suggests that
for most practical purposes, a secret key with virtually any required length $L$ can be realized by the AAA method. Note that $\epsilon_j^{(L)}$ equals $L$ whenever there is $\mu_i=1$ for some $i\in\{1,\cdots,j\}$. In practice, at time $j$, $\mu_i$ is either zero or one for $i\leq j$.

\section{Discussions}\label{sec:discussion}

\subsection{Comparison to other alternatives and an optimal method}

Given a single realization of $\mathcal{X}_j\doteq\{X_1,\cdots,X_j\}$, the AAA method produces a single bit $K_j$. Obviously, one could partition $\mathcal{X}_j$ into multiple subsets, and each of the subsets could be used in the same way to produce a bit, which results in multiple bits from the single realization of $\mathcal{X}_j$. This is a trivial variation, and the properties of the resulting key follow straightforwardly from those shown earlier.

There are clearly other possible ways one can use to map $\mathcal{X}_j$ into some part of a secret key. But whether or not there is a more effective way than the AAA method is an open question.

However,  the optimal/maximum number of (independent) secret bits the users can extract out \emph{per  independent realization} of $\mathcal{X}_j$  is given by the following equivocation:
\begin{equation}\label{eq:optimal}
  \mathbb{H}\{\mathcal{X}_j|\mathcal{E}_j\}=\sum_{i=1}^j\mathbb{H}\{X_i|E_i\}=\sum_{i=1}^j\mu_i.
\end{equation}
See \eqref{eq:E1} for the fact $\mathbb{H}\{X_i|E_i\}=\mu_i$.
This means that in theory each realization of $\mathcal{X}_j$ ``carries'' $\sum_{i=1}^j\mu_i$ secret bits against Eve.  The difference between \eqref{eq:optimal} and \eqref{eq:Ej} (with the arbitrary $l$ ignored) is an optimality gap of the AAA method, which can be significant unless $\mu_i$ for all $i$ are very small,.

Yet, the above is only part of the picture one should see. To realize the optimal limit shown in  \eqref{eq:optimal}, i.e., to produce a key of the size equal to \eqref{eq:optimal}, the users must have and apply the following assumptions:
\begin{enumerate}
   \item $\mu_1,\cdots,\mu_j$ are known to users;
  \item There is a large number of independent realizations of $\mathcal{X}_j$ (and $\mathcal{E}_j$) subject to the same $\mu_1,\cdots,\mu_j$.
\end{enumerate}

For the intended applications of the AAA method, none of the above assumptions is feasible. As discussed earlier, the bit stream $X_1, \cdots, X_j$ of interest is generally non-stationary; and multiple ``realizations'' of $X_i$,  chosen from file $i$ which is subject to the probability $1-\mu_i$ of intercept by Eve, are generally not independent of each other when conditioned on Eve's observations.

If $\mu_1,\cdots,\mu_j$ (probabilities of miss at Eve) are replaced by their estimates or some (``conservative'') lower bounds $\mu_1',\cdots,\mu_j'$, then the number of secret-key bits achieved by an ``optimal'' method would be $\sum_{i=1}^j\mu_i'$, which could be smaller than $\epsilon_j$ in \eqref{eq:Ej}. For example, if the true values of $\mu_1,\cdots,\mu_j$ at time $j$ are $0,1,0,\cdots,0$, and the chosen estimate (used by the ``optimal'' method) for $\mu_1,\cdots,\mu_j$  is a small number $\mu<\frac{1}{j}$, then $\epsilon_j=1$ and $\sum_{i=1}^j\mu_i'=\mu j<\epsilon_j$.

Furthermore, for the intended applications, the true value of $\mu_i$ at any time $j\geq i$ is either zero or one (i.e., not a number between zero and one). This is because at the time $j\geq i$, the data file associated with $X_i$ is either intercepted or not intercepted.  To users, $\mu_i$ for every $i$ is a binary unknown number. The lower and upper bounds on $\mu_i$ would be respectively zero and one, which are not useful for any optimal method. (Strictly speaking, $\mu_i$ not only depends on  the time $i$ when $X_i$ is transmitted between users but also other possible events before the current time $j$.)

  \subsubsection{An optimal method}
To better appreciate the simplicity and practical importance of the AAA method, let us now look deeper into how the capacity in \eqref{eq:optimal} can be typically realized by an optimal method.
First, note that the data set $\mathcal{X}_j$ is available at all users, and hence there is no error-correction issue here but only the privacy amplification issue.

Let $n$ independent realizations of $\mathcal{X}_j=\{X_1,\cdots,X_j\}$
 be denoted by $\mathcal{X}_j^{(m)}=\{X_1^{(m)},\cdots,X_j^{(m)}\}$
   with $m=1,\cdots,n$. Then each user can apply an identical hash function $H_i$ (a randomly chosen linear function in the binary field) depending on  $i$  to compress $X_i^{(1)}, \cdots, X_i^{(n)}$ into $\lfloor\mu_in\rfloor$ i.i.d. uniformly random bits with $i=1,\cdots,j$. (For a large $n$, and conditioned on Eve's observations,  $\mu_i n$ entries/bits of $X_i^{(1)}, \cdots, X_i^{(n)}$ are i.i.d. uniformly random, and the other $(1-\mu_i)n$ entries/bits are constant. Hence, conditioned on Eve's observations, a typical linear function $H_i$  provides a one-to-one function between the $\mu_i n$ bits missed by Eve and the $\mu_i n$ output bits.) Hence,  for a large $n$, i.e., $n\to\infty$, the capacity in \eqref{eq:optimal}  is realized.
Once again, the optimal method requires the use of $\mu_i$ for all $i$, as well as multiple independent realizations of $\mathcal{X}_j$ (and $\mathcal{E}_j$).

\subsection{Connection with Maurer's theory}

If there are $n$ independent realizations of $\mathcal{X}_j\doteq\{X_1,\cdots,X_j\}$ and $\mathcal{E}_j\doteq\{E_1, \cdots, E_j\}$, and $\mu_i$ for all $i=1,\cdots,j$ are publicly known, then it has been established in \cite{Bloch2011}, \cite{Maurer1993}, and \cite{Ahlswede1993} that the secret-key capacity $C_{\texttt{key}}$ (in bits per independent realization of $\mathcal{X}_j$ and $\mathcal{E}_j$), achievable via public communications and as $n\to\infty$, is lower bounded by $l\doteq \mathbb{I}(\mathcal{X}_j;\mathcal{X}_j)-\mathbb{I}(\mathcal{X}_j;\mathcal{E}_j)$ and upper bounded by $u\doteq \mathbb{I}(\mathcal{X}_j;\mathcal{X}_j|\mathcal{E}_j)$. Here $\mathbb{I}(A;B|C)$ denotes the mutual information between $A$ and $B$ given $C$. Furthermore, one can verify that $C_{\texttt{key}}=l=u=\sum_{i=1}^j \mu_i$
which is the same as \eqref{eq:optimal}.
Specifically, $l=\sum_{i=1}^j (\mathbb{I}(X_i;X_i)-\mathbb{I}(X_i;E_i))
=\sum_{i=1}^j([\mathbb{H}(X_i)-\mathbb{H}(X_i|X_i)]-[\mathbb{H}(X_i)-\mathbb{H}(X_i|E_i)])
=\sum_{i=1}^j\mathbb{H}(X_i|E_i)=\sum_{i=1}^j \mu_i$, and $u=\sum_{i=1}^j \mathbb{I}(X_i;X_i|E_i)=\sum_{i=1}^j(\mathbb{H}(X_i|E_i)-\mathbb{H}(X_i|X_i,E_i))=\sum_{i=1}^j\mathbb{H}(X_i|E_i)=\sum_{i=1}^j \mu_i$.

\subsection{Connection with Wyner's theory}
The works by Wyner and many others on wiretap channel (WTC) systems (e.g., see \cite{Bloch2011}, \cite{Poor2017} and \cite{Wyner1975}) address how to encode and decode an individual packet (with a sufficient length over a statistically stationary channel) so that the secret information in the packet can be securely transmitted from Alice to Bob against Eve. The achievable secrecy rate of WTC in bits per symbol interval is known to be $\max_{p_X}(\mathbb{I}(X;Y)-\mathbb{I}(X;Z))$ where $p_X$ is the distribution of the symbol $X$ transmitted from Alice, and $Y$ and $Z$ are the symbols received by Bob and Eve respectively.

The AAA method shown in this paper is not about any individual packet. But instead it exploits the abundance of packets that are continuously exchanged between two or more nodes over dynamic networks where not all of these packets are likely to be intercepted by Eve. Such an example could be two or more users who are located in different cities around the world and continuously exchange many packets with each other over different wireless network carriers, different software applications, and/or different network servers. Unless all the carriers and servers store all packets transmitted over their networks and software platforms, and also collude with each other (which is highly impractical), a secret key (with growing secrecy until it is perfect) between the users can be established by the AAA method.

However, one can also think of an abstract WTC scheme as follows. Let Alice and Bob have the set of bits in $\mathcal{X}_j$ taken from their previously exchanged packets, and $\mathcal{E}_j$ be the corresponding set of bits at Eve. Then Alice could transmit a set of uniform random binary bits $\mathcal{S}_j\doteq \{S_1,\cdots,S_j\}$ to Bob by publicly transmitting $\mathcal{S}_j\oplus\mathcal{X}_j$ (element-wise Xor).   In this case, one can verify that the achievable secrecy rate from Alice to Bob in bits per realization of $\mathcal{X}_j$, $\mathcal{S}_j$ and $\mathcal{E}_j$, i.e., $\mathbb{I}(\mathcal{S}_j;\mathcal{X}_j,\mathcal{S}_j\oplus \mathcal{X}_j)-\mathbb{I}(\mathcal{S}_j;\mathcal{E}_j,\mathcal{S}_j\oplus \mathcal{X}_j)$ \cite{Bloch2011}, is also given by \eqref{eq:optimal}. Again, this secrecy is only achievable asymptotically over a \emph{large} number of realizations of $\mathcal{X}_j$ and with the knowledge of $\mu_i$ for all $i$.

Alternatively, one can think of another WTC where $X_1,\cdots,X_j$ are the symbols transmitted by Alice and received by Bob, and $E_1,\cdots,E_j$ are the symbols received by Eve. In this case, the secrecy capacity of the WTC in bits per $j$ symbol intervals is known to be $\sum_{i=1}^j(\mathbb{I}(X_i;X_i)-\mathbb{I}(X_i;E_i))=\sum_{i=1}^j\mu_i$, which is again equal to \eqref{eq:optimal}.
But this scheme requires a joint coding over $X_1,\cdots,X_j$, and assumes a large $j$ and $\mu_i$ being not only \emph{invariant} to $i$ but also known to the transmitter of $X_1,\cdots,X_j$ before the transmission begins. None of these assumptions is feasible in the context of interest in this paper.

\subsection{Effect of correlation or dependency}\label{sec:correlation}

In the analyses shown so far, it has been assumed that the entries in $\mathcal{X}_j=\{X_1,\cdots,X_j\}$ are independent of each other, which is needed for the best performance and can be realized rather easily in practice. With a single realization of $\mathcal{X}_j=\{X_1,\cdots,X_j\}$ and $\mathcal{E}_j=\{E_1,\cdots,E_j\}$, it is generally infeasible for Eve to extract the knowledge of dependency (if any) among $X_1,\cdots,X_j$. Nevertheless, it is shown next that the AAA method also has a robustness against such dependency even if it is known to Eve.

Let us now assume that the binary entries in $\mathcal{X}_j$ form a Markov chain with
$p(x_i|x_{i-1},\cdots,x_1)= p(x_i|x_{i-1})$, $p(x_i=x_{i-1}|x_{i-1})=\alpha_i$, $p(x_i\neq x_{i-1}|x_{i-1})=1-\alpha_i$,
and $p(x_1)=\frac{1}{2}$.  A correlation exists if $\alpha_i\neq \frac{1}{2}$ A weak correlation or dependency exists if $\alpha_i$ is in a neighborhood around $\frac{1}{2}$. It follows by symmetry that $p(x_i)=\frac{1}{2}$ for $x_i\in\{0,1\}$ and all $i\geq 1$. Also $p(x_{i-1}|x_i)=\frac{p(x_i|x_{i-1})p(x_{i-1})}{p(x_i)}=p(x_i|x_{i-1})$.
The model for $\mathcal{E}_j=\{E_1,\cdots,E_j\}$ is the same as before.

The equivocation of $K_j$ (an arbitrary bit in a secret key) to Eve is the same as defined before for $\epsilon_{l,j}$ with $l$ removed. As shown in \eqref{eq:E1}, we have $\epsilon_1= \mu_1$. Furthermore,
\begin{align}\label{eq:epsilon_2}
&\epsilon_2 = - \sum_{k_2,\mathbf{e}_2}p(\mathbf{e}_2)p(k_2|\mathbf{e}_2)\log p(k_2|\mathbf{e}_2)
\end{align}
where the distributions of $p(k_2|\mathbf{e}_2)$ and $p(\mathbf{e}_2)$ are shown below:
\begin{equation}\label{Tabel_2}
  \begin{array}{c|c|c|c|c}
    p(k_2|\mathbf{e}_2) & k_2\in\{0,1\} & e_1&e_2 & p(\mathbf{e}_2) \\
    \hline
    1 & x_1\oplus x_2 & x_1&x_2 & (1-\mu_1)(1-\mu_2)p(x_1,x_2) \\
    \alpha_2, 1-\alpha_2 & x_2,\bar x_2 & \dagger &x_2 & \mu_1(1-\mu_2)p(x_2) \\
    \alpha_2,1-\alpha_2 & x_1,\bar x_1 & x_1&\dagger & (1-\mu_1)\mu_2p(x_1) \\
    \frac{1}{2},\frac{1}{2} & 0,1 & \dagger&\dagger & \mu_1\mu_2
  \end{array}
\end{equation}
Here $\bar x\doteq x\oplus 1$, $k_2=x_1\oplus x_2$ and $\mathbf{e}_2=(e_1,e_2)$.   Substituting \eqref{Tabel_2} into \eqref{eq:epsilon_2} yields
\begin{equation}\label{eq:epsilon_2_2}
  \epsilon_2 = [\mu_1(1-\mu_2)+(1-\mu_2)\mu_1]h(\alpha_2)+\mu_1\mu_2,
\end{equation}
where $h(\alpha)\doteq -\alpha\log\alpha-(1-\alpha)\log (1-\alpha)$.

It is easy to verify that $\epsilon_2$ in \eqref{eq:epsilon_2_2} is upper bounded by $\epsilon_2$ in \eqref{eq:Ej}, with equality if $\alpha_2=\frac{1}{2}$ or equivalently $h(\alpha_2)=1$. This is an example showing that correlations in $\mathcal{X}_j$ in general reduce the equivocation.

Also, if $\alpha_2$ equals zero or one, then $\epsilon_2$ in \eqref{eq:epsilon_2_2} reduces to $\mu_1\mu_2$, which is smaller than $\epsilon_1=\mu_1$. This is an example showing that (strong) correlations in $\mathcal{X}_j$ could make $\epsilon_j$ a decreasing function of $j$. However, if $h(\alpha_2)>\frac{1}{2}$ (a ``weak'' correlation between $X_1$ and $X_2$), then $\epsilon_2>\epsilon_1$ subject to $\mu_1=\mu_2<1$.

To consider $\epsilon_3$, one can verify the following table for $p(k_3|\mathbf{e}_3)$ and $p(\mathbf{e}_3)$:
\begin{equation}\label{Tabel_3}
  \begin{array}{c|c|c|c|c|c}
    p(k_3|\mathbf{e}_3) & k_3\in\{0,1\} & e_1 & e_2 & e_3 & p(\mathbf{e}_3) \\
    \hline
    1 & x_1\oplus x_2\oplus x_3 & x_1 & x_2 & x_3 & (1-\mu_1)(1-\mu_2)(1-\mu_3)p(x_1,x_2,x_3) \\
    \alpha_2,1-\alpha_2 & x_2\oplus x_3,\overline{x_2\oplus x_3} & \dagger & x_2 & x_3 &
    \mu_1(1-\mu_2)(1-\mu_3)p(x_2,x_3) \\
    p_a^*,1-p_a^* & x_1\oplus x_3,\overline{x_1\oplus x_3} & x_1 & \dagger & x_3 & (1-\mu_1)\mu_2(1-\mu_3)p(x_1,x_3) \\
    \alpha_3,1-\alpha_3 & x_1\oplus x_2,\overline{x_1\oplus x_2} & x_1 & x_2 & \dagger & (1-\mu_1)(1-\mu_2)\mu_3p(x_1,x_2) \\
    p_b,1-p_b & x_3,\bar x_3 & \dagger & \dagger & x_3 & \mu_1\mu_2(1-\mu_3)p(x_3) \\
     p_c,1-p_c& x_2,\bar x_2 & \dagger & x_2 & \dagger & \mu_1(1-\mu_2)\mu_3 p(x_2)\\
    p_d,1-p_d&x_1,\bar x_1&x_1&\dagger&\dagger&(1-\mu_1)\mu_2\mu_3p(x_1)\\
    \frac{1}{2},\frac{1}{2} & 0,1 & \dagger & \dagger& \dagger & \mu_1\mu_2\mu_3
  \end{array}
\end{equation}
where the probabilities $p_a^*$, $p_b$, $p_c$ and $p_d$ are shown next. Note that for example, when $x_2$ and $x_3$ are observed by Eve (the 3rd line in the table), $k_3$ is either $x_2\oplus x_3$ or its complement $\overline{x_2\oplus x_3}$, and its probability conditioned on $\mathbf{e}_3$ is equal to either $\alpha_2$ or $1-\alpha_2$ (with no particular order).

To derive $p_a^*$ in \eqref{Tabel_3}, let us consider (also applying the Markov model)
\begin{align}
&p(x_2|x_1,x_3)
=\frac{p(x_1,x_2,x_3)}{\sum_{x_2}p(x_1,x_2,x_3)}
=\frac{p(x_2|x_1)p(x_3|x_2)}{\sum_{x_2}p(x_2|x_1)p(x_3|x_2)}.
\end{align}
Note that the outcome of $x_2$ is affected by both $x_1$ and $x_3$.
Then
$p_a^*\doteq p[k_3=x_1\oplus x_3|\mathbf{e}_3=(x_1,\dagger,x_3)]= p(x_2=0|x_1,x_3)$ which has four possible values, i.e., for $(x_1,x_3)=(0,0),(1,1),(1,0),(0,1)$ respectively,
$p_a^*=\frac{\alpha_2\alpha_3}{\alpha_2\alpha_3+(1-\alpha_2)(1-\alpha_3)},
\frac{(1-\alpha_2)(1-\alpha_3)}{\alpha_2\alpha_3+(1-\alpha_2)(1-\alpha_3)},
\frac{(1-\alpha_2)\alpha_3}{\alpha_2(1-\alpha_3)+(1-\alpha_2)\alpha_3},
\frac{\alpha_2(1-\alpha_3)}{(1-\alpha_2)\alpha_3+\alpha_2(1-\alpha_3)}$
.

Note that the above case, where $x_1$ and $x_3$ (but not $x_2$) have been observed by Eve, creates total 8 distinct terms in $\epsilon_3$.
Also note that since $p_a^*$ is not invariant to $(x_1,x_3)$, then the explicit expression of $p(x_1,x_3)$ in \eqref{Tabel_3} is needed, which can be shown to be
\begin{align}\label{}
  &p(x_1,x_3)=\frac{1}{2}\sum_{x_2}p(x_2|x_1)p(x_3|x_2)=\left \{\begin{array}{cc}
             \frac{1}{2}(\alpha_2\alpha_3+(1-\alpha_2)(1-\alpha_3)), & (x_1,x_3)=(0,0),(1,1); \\
             \frac{1}{2}((1-\alpha_2)\alpha_3+\alpha_2(1-\alpha_3)), & (x_1,x_3)=(1,0),(0,1) .
           \end{array}
  \right .\notag
\end{align}

To derive $p_b$ in \eqref{Tabel_3}, first consider
$
  p(x_1,x_2|x_3)=\frac{p(x_1,x_2,x_3)}{p(x_3)}=p(x_2|x_1)p(x_3|x_2)
$.
Then
$p_b\doteq p(k_3=x_3|x_3)=p(x_1=x_3, x_2=x_3|x_3)+p(x_1=x_2, x_2\neq x_3|x_3)=\alpha_2\alpha_3+\alpha_2(1-\alpha_3)=\alpha_2
$

To derive $p_c$ in \eqref{Tabel_3}, consider
$
  p(x_1,x_3|x_2) = \frac{p(x_1,x_2,x_3)}{p(x_2)}=p(x_2|x_1)p(x_3|x_2)
$.
Then
$p_c\doteq p(k_3=x_2|x_2)=p(x_1=x_2, x_3=x_2|x_2)+p(x_1\neq x_2, x_3\neq x_2|x_2)=\alpha_2\alpha_3+(1-\alpha_2)(1-\alpha_3)
$.

Similarly, $p_d$ in \eqref{Tabel_3} is
$p_d\doteq p(k_3=x_1|x_1)=p(x_2=x_1,x_3=x_2|x_1)+p(x_2\neq x_1,x_3=x_2|x_1)=\alpha_2\alpha_3+(1-\alpha_2)\alpha_3=\alpha_3
$.

Applying \eqref{Tabel_3} to $\epsilon_3$ in \eqref{eq:Hkjzj} with $j=3$, one can obtain a closed form expression of $\epsilon_3$.  One can verify that if $\alpha_2=\alpha_3=\frac{1}{2}$ and hence $p_a^*=p_b=p_c=p_d=\frac{1}{2}$, then this $\epsilon_3$ is the same as that given by \eqref{eq:Ej} with $j=3$. On the other hand, if both $\alpha_2$ and $\alpha_3$ are either zero or one (extreme correlation), and hence so are $p_a^*$, $p_b$, $p_c$ and $p_d$, then this $\epsilon_3$ becomes $\mu_1\mu_2\mu_3$.

 If $\mu_i=\mu$ and $\alpha_i=\alpha$, and let $\mu'=1-\mu$, $\alpha'=1-\alpha$, $p_a=\frac{\alpha^2}{\alpha^2+\alpha'^2}$ and $p_c=\alpha^2+\alpha'^2$ ($\geq \frac{1}{2}$), then one can verify
\begin{equation}\label{}
  \epsilon_3=\mu^3+2\mu\mu'h(\alpha)+\mu^2\mu'h(p_c)
  +\mu\mu'^2[2\alpha\alpha'+(\alpha^2+\alpha'^2)h(p_a)].
\end{equation}
Furthermore, let $\eta\doteq 2\alpha\alpha'+(\alpha^2+\alpha'^2)h(p_a)$, one can verify that $\epsilon_3>\epsilon_2$ if
$
  \hat\mu(\alpha) \doteq\frac{\eta}{1+\eta
  -h(p_c)}>\mu
$ (a ``weak'' correlation among $X_1,X_2,X_3$ although depending on $\mu$).
One can also verify that $\hat\mu(0)=0$, $\hat\mu(\frac{1}{2})=1$, $\hat \mu(\alpha)=\hat \mu (1-\alpha)$, and $\frac{\partial}{\partial \alpha}\hat\mu(\alpha)>0$ for $0<\alpha<\frac{1}{2}$.

However, finding a closed form expression of $\epsilon_j$ for an arbitrary $j$, even if  $\mu_i=\mu$ ($\neq 0,1$) and $\alpha_i=\alpha$ ($\neq \frac{1}{2},0,1$)  for all $i$, remains an open problem and appears hard.


\section{Conclusion}
A simple method for secret-key generation has been presented for mobile users who already have reliable and authenticated connections but want to establish their own secret keys for privacy purposes. This method is accumulative, adaptable and additive (AAA), which does not require channel reciprocity and/or noises at the physical layer, but rather exploits imperfect intercepts by Eve especially at the application layer of mobile communication networks. The AAA method applies a continuous (or successive) privacy amplification. The equivocation to Eve of each key bit generated by the AAA method using a generally non-stationary  sequence of independent bits has been shown to be always non-decreasing with time (subject to any given probabilities of intercepts at Eve) and typically increasing with time until it becomes perfect. The AAA method does not require the knowledge of probabilities of miss at Eve while any capacity-achieving method does. The AAA method does not require multiple independent realizations of a non-stationary data process while any capacity-achieving method does.  The AAA method has also been shown to be robust against some bit correlation known to Eve although a more general theory remains open. Given the simplicity and versatility of the AAA method and the availability of the massive amount of independent bits exchanged between modern devices,  it should be an important topic to explore how the AAA method can be used to improve the security protocols for next-generation wireless networks at all layers.

\end{document}